\documentclass[twocolumn,superscriptaddress,amsmath,amssymb,showpacs,floatfix,preprintnumbers]{revtex4-1}

\usepackage{amsmath}
\usepackage{amssymb}
\usepackage{graphicx}
\usepackage{color}
\usepackage[colorlinks,citecolor=blue,linkcolor=blue]{hyperref}
\usepackage{epsfig}
\usepackage{amsmath}
\usepackage{amssymb}
\usepackage{epstopdf}
\usepackage{sidecap}
\usepackage{booktabs}
\usepackage{rotating}

\sidecaptionvpos{figure}{c}

\DeclareGraphicsExtensions{.png .jpg .pdf}

\renewcommand{\i}{\mathrm{i}\,}
\newcommand{\e}{\mathrm{e}}
\newcommand{\p}{\partial}
\newcommand{\mc}{\mathcal}
\newcommand{\mb}{\mathbf}

\newcommand{\ez}{\hat{\mb{z}}}

\newcommand{\np}{\textit{n-p}}

\begin{document}

\title{Electronic Mach-Zehnder interference in a bipolar hybrid 
monolayer-bilayer graphene junction}

\author{M. Mirzakhani}\email{mirzakhani@ibs.re.kr}
\affiliation{Center for Theoretical Physics of Complex Systems, 
Institute for Basic Science, Daejeon, 34126, South Korea}
\author{N. Myoung}\email{nmyoung@chosun.ac.kr}
\affiliation{Department of Physics Education, Chosun University, Gwangju 61452, 
Republic of Korea}
\author{F. M. Peeters}\email{francois.peeters@uantwerpen.be}
\affiliation{Department of Physics, University of Antwerp, 
Groenenborgerlaan 171, B-2020 Antwerp, Belgium}
\author{H. C. Park}\email{hcpark@ibs.re.kr}
\affiliation{Center for Theoretical Physics of Complex Systems, 
Institute for Basic Science, Daejeon, 34126, South Korea}

\date{February 23, 2021}

\begin{abstract}
Graphene matter in a strong magnetic field, realizing one-dimensional quantum 
Hall channels, provides a unique platform for studying electron interference.
Here, using the Landauer-B\"uttiker formalism along with the tight-binding model, 
we investigate the quantum Hall (QH) effects in unipolar and bipolar monolayer-bilayer 
graphene (MLG-BLG) junctions.
We find that a Hall bar made of an armchair MLG-BLG junction in the bipolar regime
results in valley-polarized 
edge-channel interferences and can operate a fully tunable Mach-Zehnder (MZ) 
interferometer device.
Investigation of the bar-width and magnetic-field dependence of the conductance oscillations
shows that the MZ interference in such structures can be drastically affected
by the type of (zigzag) edge termination of the second layer in the BLG 
region [composed of vertical dimer or non-dimer atoms].
Our findings reveal that both interfaces exhibit a double set of Aharonov-Bohm 
interferences, with the one between two oppositely valley-polarized edge channels
dominating and causing a large-amplitude conductance oscillation ranging from 0
to $ 2 e^2 / h$. 
We explain and analyze our findings by analytically solving the Dirac-Weyl equation 
for a gated semi-infinite MLG-BLG junction.
\end{abstract}

\maketitle


\section{Introduction} \label{intro}

Bipolar (\np) graphene junctions, which form two neighboring regions with different 
quantum Hall (QH) states, are of tremendous interest for low-dimensional materials 
because they exhibit unusual and fascinating transport properties
\cite{Williams2007,Abanin2007,Huard2007,Ozyilmaz2007}.
For instance, Abanin et al.~\cite{Abanin2007} first theoretically addressed QH transport
for two regions with different filling factors ($\nu_1$ and $\nu_2$),
predicting new quantized conductance plateaus at values of 
$G = \frac{|\nu_1||\nu_2|}{|\nu_1| + |\nu_2|} \frac{e^2}{h}$.
This prediction was verified by Williams et al.~\cite{Williams2007} 
in experimental transport measurements of the MLG \np\ junction. 
Theoretical studies \cite{Tworzydlo2007,Myoung2017,Trifunovic2019,Myoung2020} 
showed that valley-isospin conservation plays an important role in the evolution
of edge states into interface states along  
the \np\ junction of MLG nanoribbons, confirming valley isospin conservation in
ribbons with armchair boundaries \cite{Tworzydlo2007,Trifunovic2019}.
Recent experimental studies also show that graphene \np\ junction at high 
magnetic fields hosts (valley- and spin-polarized) edge channels propagating 
along the junction where coupling between those channels results in a 
Mach-Zehnder (MZ) interferometer \cite{Ji2003} showing Aharonov-Bohm (AB) effect
\cite{Morikawa2015,Wei2017,Makk2018,Jo2021}. 

In addition to bipolar junctions of graphene, hybrid graphene structures 
consisting of two different areas of graphene layers, e.g., partly monolayer 
graphene (MLG) and partly bilayer graphene (BLG), 
also present a lot of interesting physics
\cite{Puls2009,Tian2013,Yan2016,Zhao2016,Du2021}.
The Hall resistance across such graphene hybrid structures shows quantized plateaus 
switching between
those of MLG or BLG QH plateaus depending on the type of carriers 
\cite{Tian2013,Yan2016}.
Experimental study on a dual-gated MLG-BLG junction recently revealed that such 
graphene channels exhibit different conductance under bias voltage in opposite 
directions, which also depends on the doping level \cite{Du2021}.
The energy spectrum of a semi-infinite MLG-BLG junction in the presence of a magnetic 
field was theoretically studied in Ref.\ \cite{Koshino2010}. 
The results showed that the valley degeneracy is lifted near the MLG-BLG interface, 
and oscillatory band structures appear in the boundary region.
Such interface states were previously observed as \textit{anomalous} resistance
oscillations in an experimental work by Puls \textit{et al} \cite{Puls2009}.
The observation of interface states in a \textit{natural} junction between MLG and BLG
suggests that this graphene system might be used to explore QH edge interface and 
electron interference in hybrid structures comprising various LL configurations.
Electron interferometry is one of the most promising routes 
for studying coherence 
effects of electronic states \cite{Litvin2007,Bieri2009,Bocquillon2013},
noise in collision experiments \cite{Henny1999,Oliver1999},
fractional and non-Abelian statistics \cite{Law2006,Feldman2006,Bid2009},
and quantum entanglement via two particle interference 
\cite{Yurke1992,Samuelsson2004}. 

\begin{figure}
	\centering
	\includegraphics[width = 8.5 cm]{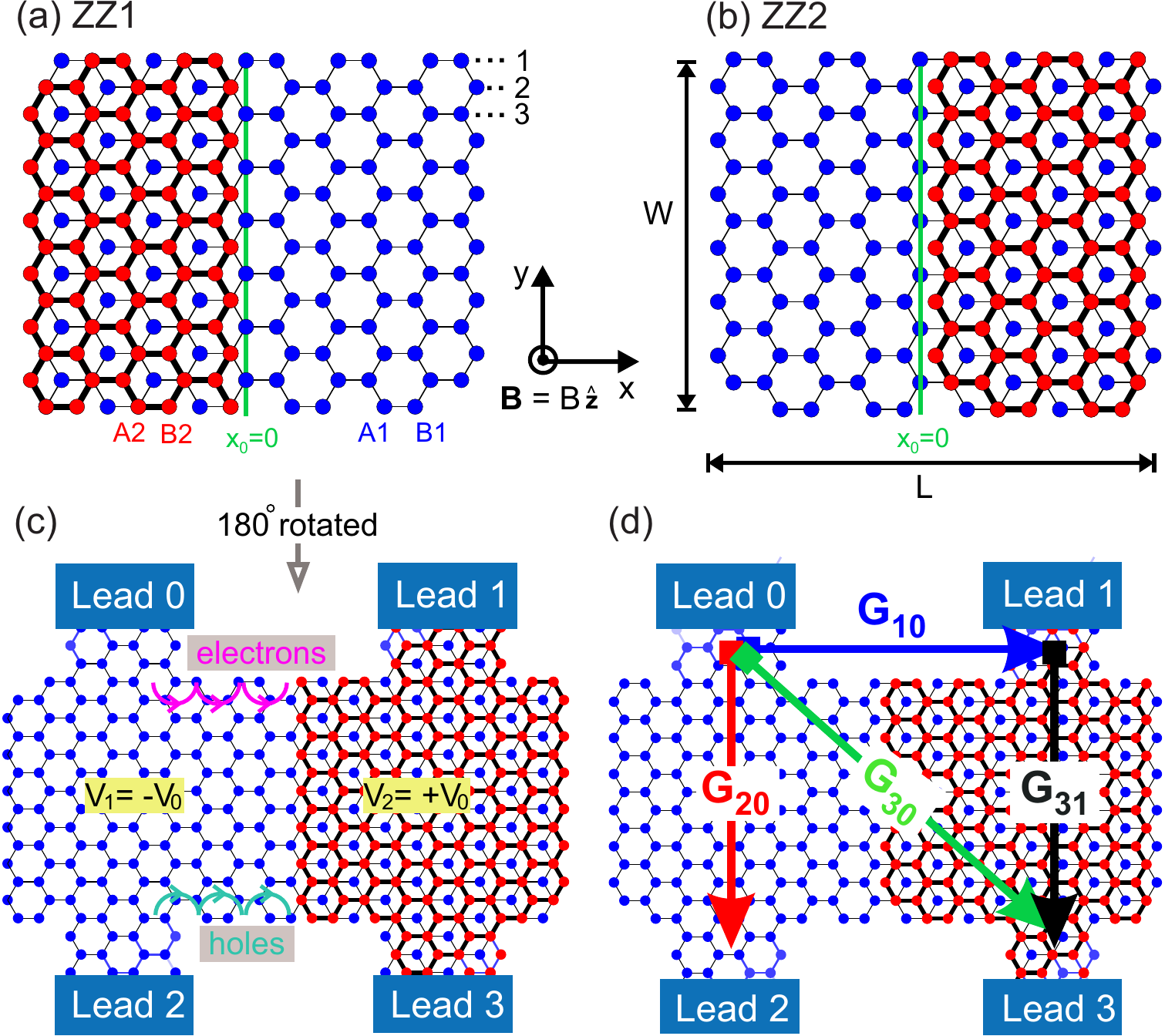}
	\caption{Schematics of a MLG-BLG Hall bar structure (width $W$ and length $L$)
	with (a) Zigzag 1 (ZZ1) and (b) Zigzag 2 (ZZ2) junctions. 	
	(c,d) Four-terminal depiction of the above junctions and the corresponding 
	numbering used to refer to the leads. 
	To describe both ribbons with similar geometry, we rotate panel (a) by 
	$180^\circ$	to obtain panel (c).
	The purple and cyan snake-like curves in panel (c), respectively, 
	indicate the electron and hole
	edge state chiralities for the given direction of the magnetic field, $\ez$ axis.	
	Bipolar MLG-BLG junction is created by applying the potentials 
	$V_1 = -V_0$ and $V_2 = V_0$ to the MLG and BLG parts, respectively.
	The longitudinal conductance $G_{10}$ (blue) and Hall conductances of the MLG part 
	$G_{20}$ (red), BLG part $G_{31}$ (black), and MLG-BLG interface $G_{30}$ (green)
	are shown in panel (d) with the corresponding colors which we use to indicate them
	in the numerical plots. 
	}
\label{fig1}
\end{figure}

Here, we demonstrate that the \np\ junction of the MLG-BLG interface bar in a Hall
regime results in valley-polarized edge-channel interferences and can operate 
as an electronic MZ interferometer device.
In this paper, using the tight binding model (TBM), we investigate the conductance
properties of the MLG-BLG interface for both unipolar and bipolar junctions in the QH regime.
We calculate the longitudinal and Hall interface conductances for two types of 
MLG-BLG interfaces: zigzag 1 [ZZ1, Fig.\ \ref{fig1}(a)] and zigzag 2 
[ZZ2, Fig.\ \ref{fig1}(b)] using a four-terminal Hall conductor made up of partial 
MLG and BLG nanoribbons with armchair edges, as shown in Figs.~\ref{fig1}(c) 
and \ref{fig1}(d).
ZZ1 is composed of vertical \textit{dimer} sites ($B1$-$A2$), whereas ZZ2 only has 
\textit{non-dimer} $B2$ atoms.

In the case of the unipolar junction, unlike the ZZ2 interface, which exhibits 
a series of well-realized QH plateaus for longitudinal interface conductance (LIC),
the ZZ1 interface shows irregular LIC fluctuations beyond the first QH plateau.
In the bipolar regime, both types of MLG-BLG interfaces exhibit a gate tunable
Hall interface conductance (HIC) with resonant behavior as a function of Fermi
energy.
We show that these oscillations result from the AB interference between
valley-polarized QH edge channels propagating along the MLG-BLG \np\ junction.
We analyze our findings by solving the Dirac-Weyl equation analytically and 
obtaining the energy levels for an ideal \np\ junction of semi-infinite MLG-BLG 
interface.
The results show that (three) valley-polarized edge channels 
(but still spin degenerate) are formed near the MLG-BLG interface, 
along the \np\ junction, and their spatial separations are energy dependent.
The spectra of these edge channels differ for ZZ1 and ZZ2 interfaces, 
resulting in distinct conductance behaviors in each case.  

By investigating the bar-width dependence of the conductance oscillations,
we demonstrate that as a result of AB interference between those 
spatially-separated edge channels propagating along the \np\ junction, 
the cross-junction transport shows oscillatory behavior.
Our results show a (one-set) regular and a double set of AB oscillations for the
ZZ1 and ZZ2 interfaces, respectively.
However, in both cases, coupling between two opposite-valley-polarized 
edge channels is dominant, resulting in a large conductance oscillation 
with a peak-to-peak amplitude ranging between $0$ and $ 2 e^2 / h$.  
On the other hand, studying the magnetoconductances, 
we see small-amplitude conductance oscillation for both interfaces, which 
is not noticeable for the ZZ1 boundary and also suppresses as the magnetic 
field increases; 
for the ZZ2 boundary, it persists over a wide range of magnetic fields.

The rest of the paper is organized as follows: In Sec.\ \ref{theo}, we present the 
proposed structures as well as the basics of our numerical method. 
Section \ref{Nr} is dedicated to results and discussions.
In Sec.\ \ref{unipo}, we present a comprehensive study of the transport properties 
of MLG-BLG junction in the unipolar regime and its bipolar junction is 
investigated in Sec.\ \ref{bipo}.
Finally, we conclude the manuscript in Sec.\ \ref{con}.

\section{Theory and model} \label{theo}

We consider a bipolar quantum Hall graphene bar consisting of MLG-BLG junction 
as shown in Fig.\ \ref{fig1}.
Geometrically, this structure can be regarded as a (AB-stacked) 
BLG ribbon in which half 
of its upper layer is cut out, thus creating the MLG-BLG junction. 
We assume that the lower layer of BLG part, containing $A1$
and $B1$ sublattices, seamlessly continues to the MLG part with
\textit{A} and \textit{B} sublattices, while the upper graphene 
layer composed of $A2$ and $B2$ sublattices is sharply terminated 
at the boundary.
Depending on which part of the upper layer in BLG ribbon is removed, 
one would have two  distinct boundaries labeled as ZZ1 and ZZ2. 
In the case of ZZ1 termination [Fig.\ \ref{fig1}(a)], the outermost 
atoms of the upper layer are $A2$ atoms that directly couple to 
the $B1$ atoms of the lower layer (dimer atoms), whereas in the case of 
ZZ2 termination, the $B2$ atoms on the upper layer [having no counterpart from
the lower layer (non-dimer atoms)] form the front-most line of the bilayer region 
[Fig.\ \ref{fig1}(b)].
Furthermore, we use metallic armchair ribbon, the width of which is characterized 
by $N = 3m - 1$ ($m$ being an integer), referring to the number of horizontal 
dimer lines of the ribbon as illustrated in Fig.\ \ref{fig1}(a).

Using a single-orbital TBM for $p_z$ atomic orbital of carbon, which 
in a second quantization formalism can be written as
\begin{equation} \label{eq:ham}
  \mathcal{H} = \sum_i (\epsilon_i + V_i) c^\dagger_i c_i
  -\sum_{\langle i,j \rangle} t_{ij} c^\dagger_i c_j + \text{H.c.},
\end{equation}
where $c^\dagger_i$ and $c_i$ are, respectively, the creation and
annihilation operators for an electron on the $i$th lattice site with 
on-site energy $\epsilon_i$ and 
$ V(x) = V_0 \tanh(x/\xi) $ 
is a position-dependent potential applied to the structure.
As shown in Fig.\ \ref{fig1}(c), $ V(x) $ takes the values of $ -V_0 $ 
and $ +V_0 $ in the MLG ($x < 0$) and BLG ($x > 0$) regions, 
respectively, thus separating \textit{n}-doped region for $x < 0$ from 
\textit{p}-doped region for $x > 0$.
Notice that $\xi \rightarrow 0$ introduces an abrupt step \np\ junction, while 
$\xi \neq 0$ indicates a smoothly varying potential.
In this paper, we used $\xi = 0.05$~nm representing the first case. 
In the second term of Hamiltonian \eqref{eq:ham}, 
$t_{ij}$ is the hopping transfer integral between two atoms ($i$, $j$) and 
$\langle i,j \rangle$ denotes a summation over nearest neighbor sites.
Here, we use a simple model for MLG and BLG, with only 
$ \gamma_0 = -2.7 $ eV and $ \gamma_1 = 0.48 $ eV describing the nearest-neighbor 
intralayer and interlayer hopping $t_{ij}$, respectively.

The effect of a perpendicular magnetic field ($\mb{B} = B\, \ez$) can be 
introduced into the calculations via the Peierls substitution \cite{Peierls1933}
$t_{ij} \rightarrow t_{ij} \e^{\i 2\pi \Phi_{ij}}$
where
$ \Phi_{ij}= (1/\Phi_0) \int^{\mb{r}_j}_{\mb{r}_i}\mb{A}(\mb{r})\cdot d\mb{r}$
is the Peierls phase with
$\Phi_0 = h/e \approx 4.14 \times 10^{-15}\ \mathrm{T.m^2}$ 
the magnetic flux quantum and $\mb{A}(\mb{r})=(0, Bx, 0)$ the vector potential
in the Landau gauge for which $\Phi_{ij}$ is given by 
$\Phi_{ij} = (y_j - y_i)(x_j + x_i)B / 2 \Phi_0$.

In the linear response regime, the Landauer-B\"uttiker formalism provides a 
rigorous formalism to describe multi-terminal conductance measurements in
Hall bars as \cite{Datta1997}
\begin{equation} \label{eq:LB}
  G_{pq} (E) = \frac{2 e^2}{h}
  \sum_{m \in q} \sum_{n \in p} |s_{nm}|^2\,.
\end{equation}
Here, $G_{pq} (E)$ represents the conductance from lead (or terminal) $q$
to lead $p$ at the energy $E$ and 
$s_{nm}$ is the scattering matrix ($S$-matrix) from mode $m$ in lead $q$
to mode $n$ in lead $p$.
In order to investigate the conductance of the four-terminal Hall bar 
{[Figs.\ \ref{fig1}(c,d)], we use the Kwant package \cite{Kwant2014},
which employs $S$-matrix formalism in conjunction with the TBM
to calculate quantum transport properties of materials.

\begin{figure}
	\centering
	\includegraphics[width = 8 cm]{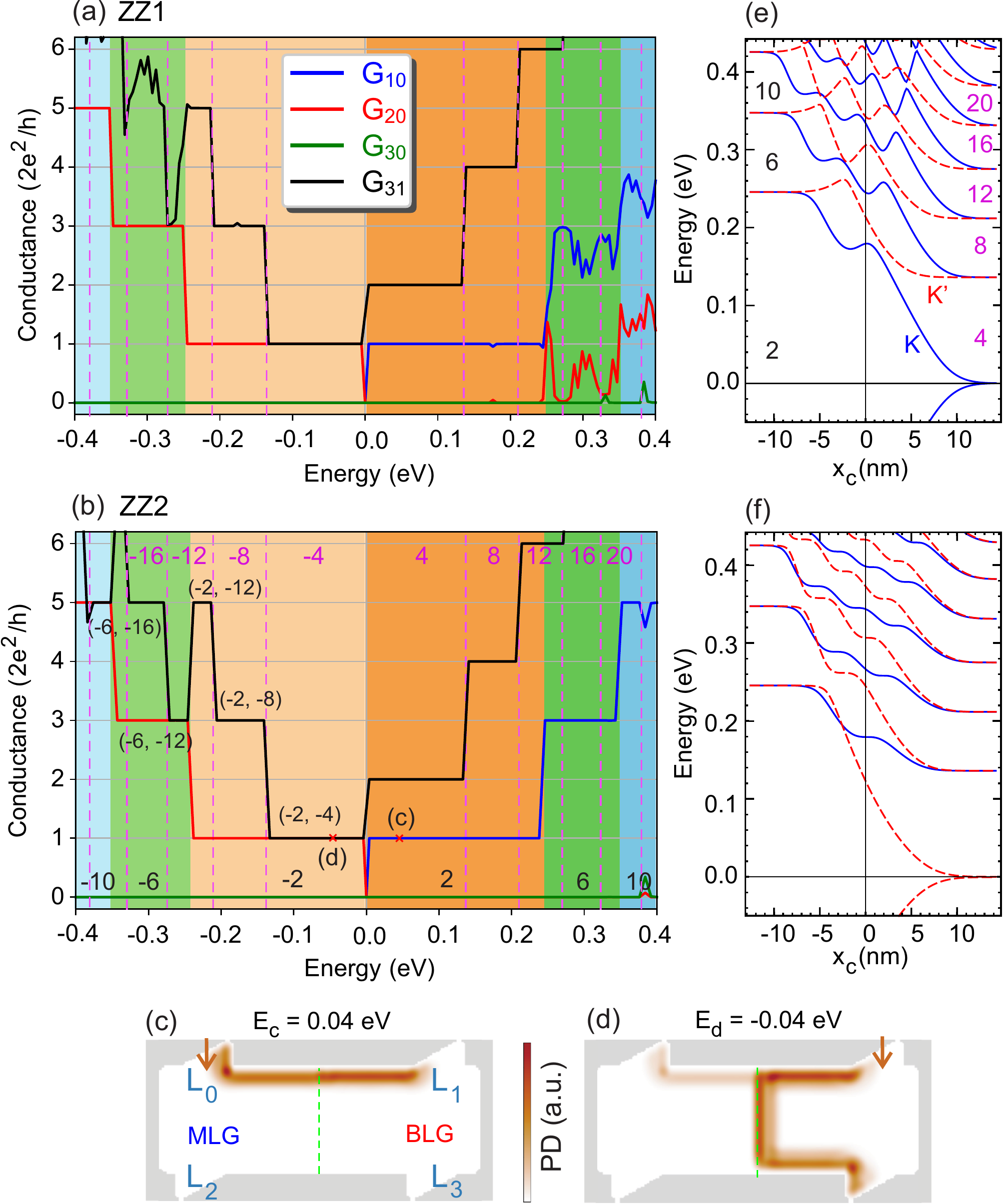}
	\caption{Four-terminal conductance of a unipolar MLG-BLG quantum Hall bar 
	as a function of the Fermi energy for (a) ZZ1 and (b) ZZ2 interfaces at the 
	magnetic field of $B = 60$ T. 
	The results are shown for a metallic armchair graphene ribbon with 
	$L = 169.87$~nm and $W = 51.91$~nm.
	Colored regions represent different bulk filling factors of MLG with
	the corresponding numbers shown in the lower part of panel (b) as 
	$\nu_1 = \pm 2, \pm 6, \pm 10$.
	The corresponding filling factors of BLG, $\nu_2 = \pm 4, \pm 8, \pm 12$, 
	are represented by the purple numbers in the upper side of the panel (b) 
	between the vertical dashed purple lines.
	(c, d) Probability densities (PD) corresponding to the energies marked by 
	(c) and (d) in panel (b) at the Fermi energies $E_c = 0.04$ and $E_d = -0.04$,
	respectively.
	In panels (c) and (d), the propagating modes enter from lead 0 and 1, 
	respectively.
	(e, f) Energy spectrum of a half-infinite MLG-BLG junction versus the 
	center of the cyclotron orbit $x_c = -k_yl_B^2$ 
	obtained by solving the Dirac-Weyl equation analytically (Appendix) 
	with (e) ZZ1 and (f) ZZ2 interfaces.
	Blue solid (red dashed) curves refer to the $K$ ($K'$) valley.
	As seen, the energy levels approach the bulk Landau quantization of MLG and BLG
	in the left and right side of the spectrum, respectively. 
	}
\label{fig2}
\end{figure}

\section{Results and discussion} \label{Nr}

\subsection{Unipolar MLG-BLG quantum Hall bar}  \label{unipo}

We first consider the unipolar and bipolar MLG-BLG Hall bar
with both ZZ1 and ZZ2 monolayer-bilayer interfaces as illustrated in Figs.\ \ref{fig1}(a)
and \ref{fig1}(b), respectively.
In the proposed four-terminal hybrid Hall bar [Figs.\ \ref{fig1}(c) and \ref{fig1}(d)], 
we measure the LIC $G_{10}$,
Hall conductances of the MLG part $G_{20}$, BLG part $G_{31}$, and MLG-BLG interface 
$G_{30}$, simultaneously.
Using these conductances enables us to individually measure the splitting of 
the conductance at the interface of the hybrid structure.
In this paper, we present the results for a metallic armchair graphene ribbon
with $L = 169.87$~nm and $W = 51.91$~nm that is subjected to a perpendicular 
magnetic field of $B=60$ T 
for which the corresponding magnetic length is $l_B = \sqrt{\hbar/e B} = 3.31$~nm.
We use a rather strong $B$ here to ensure $l_B \ll W$ as well as leads width,
$l_B < l_w \approx 25.50$~nm.
It is worth to mentioning that, for the study of the electronic properties of graphene
nanostructures in the presence of a perpendicular magnetic field,
one can define a scaling factor and thus extend the results to
lower magnetic field and larger sampsle sizes, e.g., see 
Refs.\ \cite{Makk2018,Liu2015,Cabosart2017}.
Furthermore, for the given magnetic field direction, $\ez$ axis, the corresponding 
edge state chiralities for electrons and holes are clockwise (CW) and 
anticlockwise (ACW), respectively [see Fig.\ \ref{fig1}(c)].

\begin{figure*}
	\centering
	\includegraphics[width = 17.5 cm]{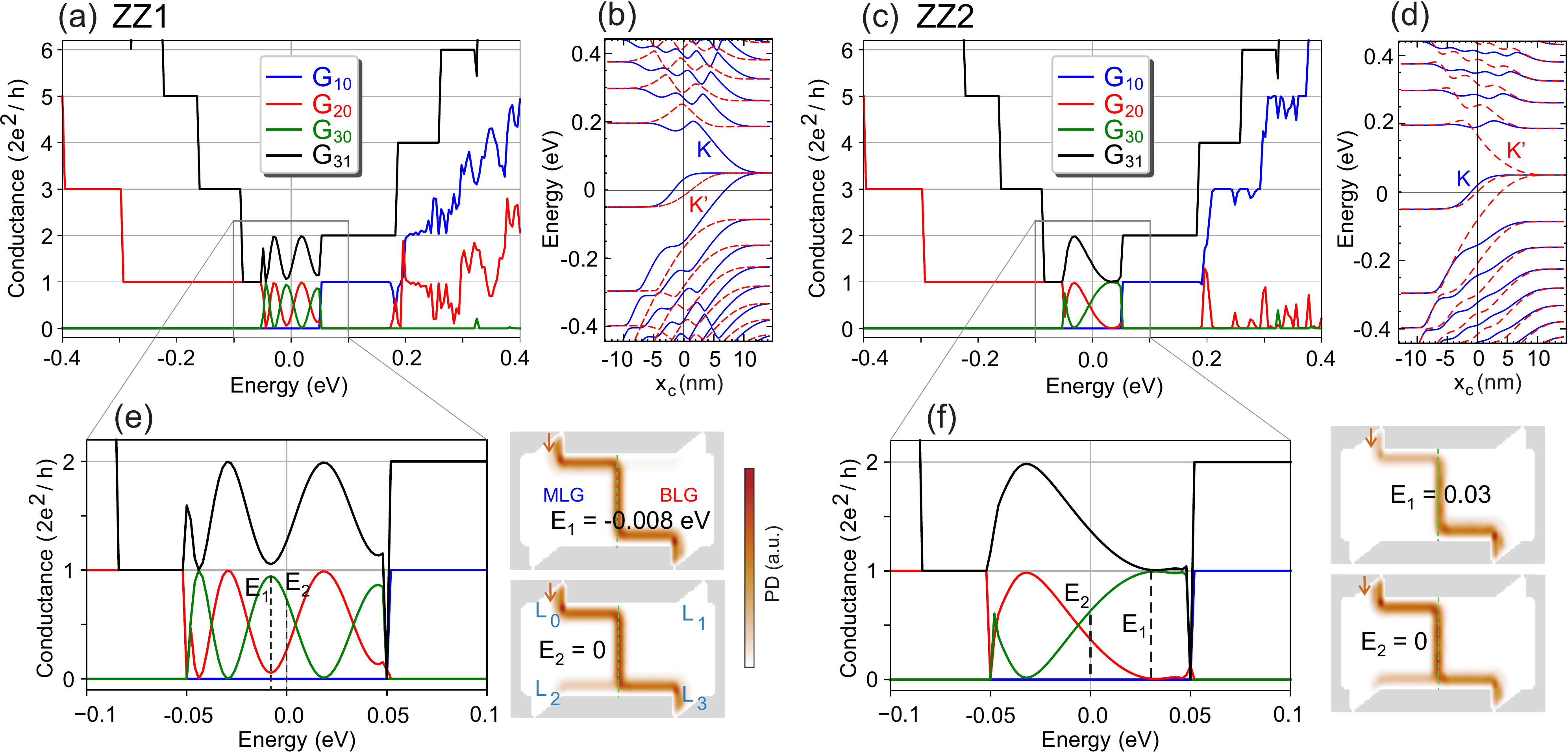}
	\caption{(a,c) Four-terminal conductance of a bipolar MLG-BLG quantum Hall bar 
	as a function of Fermi energy for (a) ZZ1 and (c) ZZ2 interfaces.
	The applied potentials to the MLG and BLG parts, respectively, are $V_1 = -V_0$
	and $V_2 = V_0$ with $V_0 = 0.05$ eV.
	(b,d) Energy spectrum of the half-infinite MLG-BLG \np\ junction versus the center 
	of the cyclotron orbit $x_c = -k_yl_B^2$ obtained analytically for
	(b) ZZ1 and (d) ZZ2 interfaces.
	Blue solid (red dashed) curves refer to the $K$ ($K'$) valley.
	(e,f) Enlarged views of the conductances around the bipolar energy range,
	as indicated by grey boxes in (a) and (c) panels.
	The rightmost figures in each panel show the probability densities (PD) of 
	the propagating modes coming in from lead 0 for each interface at the two 
	different Fermi energies $E_1$, $E_2$, which are labeled in panels (e) 
	and (f).
	}
\label{fig3}
\end{figure*}

We begin by analyzing the measured conductances $G_{10}$, $G_{20}$, $G_{30}$, 
and $G_{31}$ for a unipolar MLG-BLG junction.
Figures \ref{fig2}(a) and \ref{fig2}(b) show the conductances as a function 
of Fermi energy for (a) ZZ1
and (b) ZZ2 MLG-BLG interfaces in the unipolar junction. 
For both interfaces, we see that the $G_{20}$ (red curve) 
exhibits the standard MLG QH plateaus 
(odd numbers of $ 2e^2/h $) for hole states ($E < 0$) with the ACW chirality and the
$G_{31}$ (black curve) in the BLG part shows the BLG QH plateaus 
(even numbers of $ 2e^2/h $) for electron states ($E > 0$) with the CW chirality.
The conductance quantization represented by $G_{31}$ 
for the hole states ($E<0$) and $G_{10}$ (LIC) for $E>0$
in both ribbons can be explained by the theory
addressed in Refs.\ \cite{Williams2007,Abanin2007}.
According to this model, in a unipolar regime \textit{n-n} or \textit{p-p},
the conductance values across the interface (e.g., $G_{10}$) follow
\begin{equation}
  G = \mathrm{min} (|\nu_1|, |\nu_2|) \times \frac{e^2}{h},
\end{equation}
where $\nu_1$ ($= \pm 2,\ \pm 6,\ \ldots$) and $\nu_2$ 
($= \pm 4,\ \pm 8,\ \ldots$) are the filling factors in the MLG and BLG regions,
respectively.
As a result, the remaining edge modes $|\nu_1 - \nu_2|$ in the region of maximum
absolute filling factor propagate along the interface and return back to that region.
For example, in Fig.\ \ref{fig2}(b), each stepwise value of $G_{31}$ for hole states 
can be obtained by applying this analysis with the corresponding filling factors 
$(\nu_1,\ \nu_2)$ defined at each conductance plateau depending on the Fermi energy,
as depicted in Fig.\ \ref{fig2}(b).
Here, the colored regions represent different bulk filling
factors of MLG as $\nu_1 = \pm 2, \pm 6, \pm 10$ and the regions between the vertical 
dashed purple lines refer to the corresponding filling factors of BLG, i.e.,
$\nu_2 = \pm 4, \pm 8, \pm 12, \pm 16$.
Notice that the LLs of MLG and gapless BLG can be obtained using 
$ E_n^M = \pm v_F \sqrt{2 e \hbar n B} $
and $ E_n^B = \pm \sqrt{n (n-1)} \hbar e B  / m^* $ ($ n = 0, 1, 2, \ldots $),
respectively, with $ m^* = \gamma_1 / 2 v_F^2 $ representing  
the effective mass of quasiparticles \cite{Novoselov2005, Rutter2011}.
Further, for both ribbons, there is no HIC $G_{30}$ measured  
between leads 0 and 3 as shown by green curves in Figs.\ \ref{fig2}(a) 
and \ref{fig2}(b). 

The above discussion can be highlighted further by plotting the probability 
densities for the two representative Fermi energies denoted by (c) and (d) 
in Fig.\ \ref{fig2}(b).
For the electron Fermi energy (c) [see Fig.\ \ref{fig2}(c)], 
one can see that the coming modes from lead 0 completely pass the 
interface along the edge of the Hall bar, resulting in $G_{10} = 2 e^2 / h$.
In state (d), modes from lead 1 in the BLG region split up at the interface, 
are partially transmitted across the interface, and the remaining modes propagate 
through the interface and return to the BLG region [Fig.\ \ref{fig2}(d)].

Beyond the first MLG filling factor area, however, the longitudinal conductance 
$G_{10}$ (blue curve) and the Hall one $G_{31}$ (black curve) exhibit different 
behaviors for ZZ1 and ZZ2 interfaces.
For the ZZ2 interface, $G_{10}$ and $ G_{31} $ exhibit well-realized QH plateaus,
whereas the ZZ1 interface exhibits irregular conductance fluctuations, 
cf.\ Figs~\ref{fig2}(a) and \ref{fig2}(b).
This can be attributed to the different behavior interface states that appear 
near the boundary region.
Analytically, we solve the Dirac-Weyl equation for a composed system of a 
half-infinite graphene monolayer and bilayer, similar to the structure studied in 
Ref.\ \cite{Koshino2010} (see also the Appendix), and plot the energy levels 
as a function of the cyclotron orbit center $x_c = -k_y l_B^2$ 
($l_B$ is the magnetic length), as shown in Figs.~\ref{fig2}(e) and \ref{fig2}(f).
As seen, the interface LLs for the ZZ2 interface exhibit monotonic dependence, 
whereas it is nonmonotonic in the ZZ1, indicating that the energy coupling between 
the MLG and BLG regions is weaker in the ZZ1 interface than in the ZZ2 interface, 
as also discussed in Ref.~\cite{Koshino2010}.
Except for the fine differences reported here between the two ZZ1 and ZZ2 interfaces, 
our numerical results are consistent with the experimental results reported in 
Refs.\ \cite{Tian2013,Yan2016} for unipolar hybrid MLG-BLG structures.
Because realistic structures will have irregular and mixed edge types, 
flawless edge terminations are required to observe such differences.

\begin{figure}
	\centering
	\includegraphics[width = 7.5 cm]{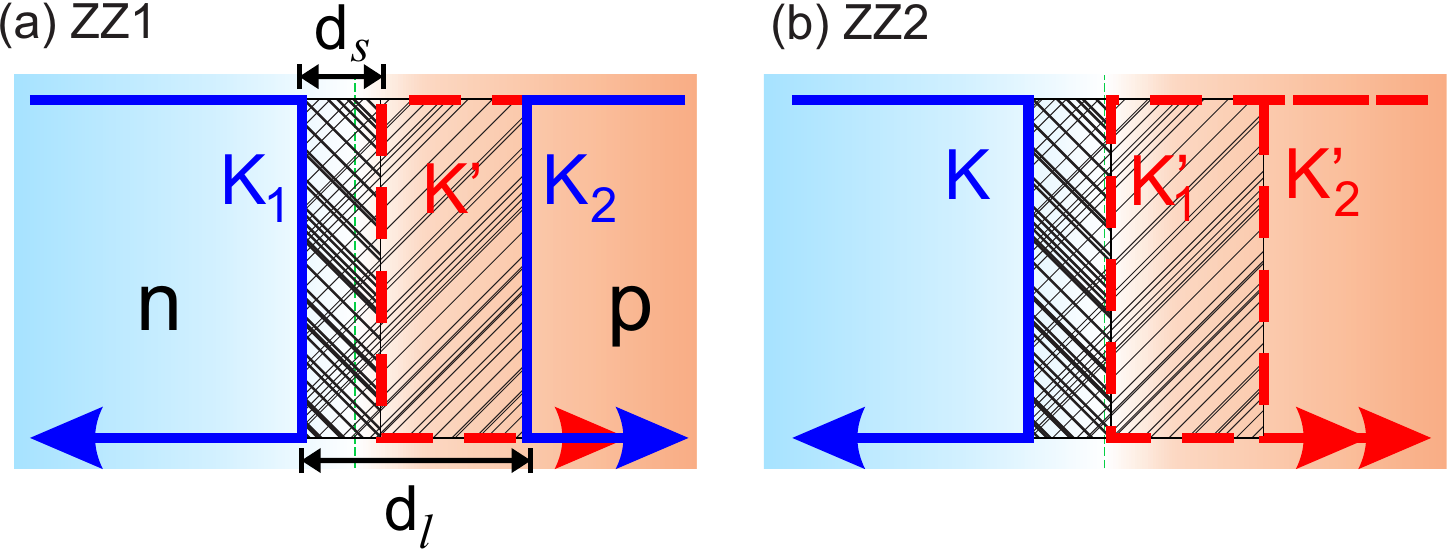}
	\caption{Schematic of edge-channel configurations and principle of AB interference
	between the edge states propagating along the \np\ junction for (a) ZZ1 and 
	(b) ZZ2 interfaces.
	The solid blue ($K$) and dashed red ($K'$) lines depict three valley-polarized 
	edge channels for each interface, as predicted by the analytical results 
	[Figs.~\ref{fig3}(b) and \ref{fig3}(d)]. 
	$d_s$ ($d_l$) shows the spatial distance between two neighboring 
	(far-distant) edge channels, and the enclosed area defined by those 
	channels is indicated by the left-angle-hatched (right-angle-hatched) area.
	}
\label{fig4}
\end{figure}

\begin{figure*}
	\centering
	\includegraphics[width = 17 cm]{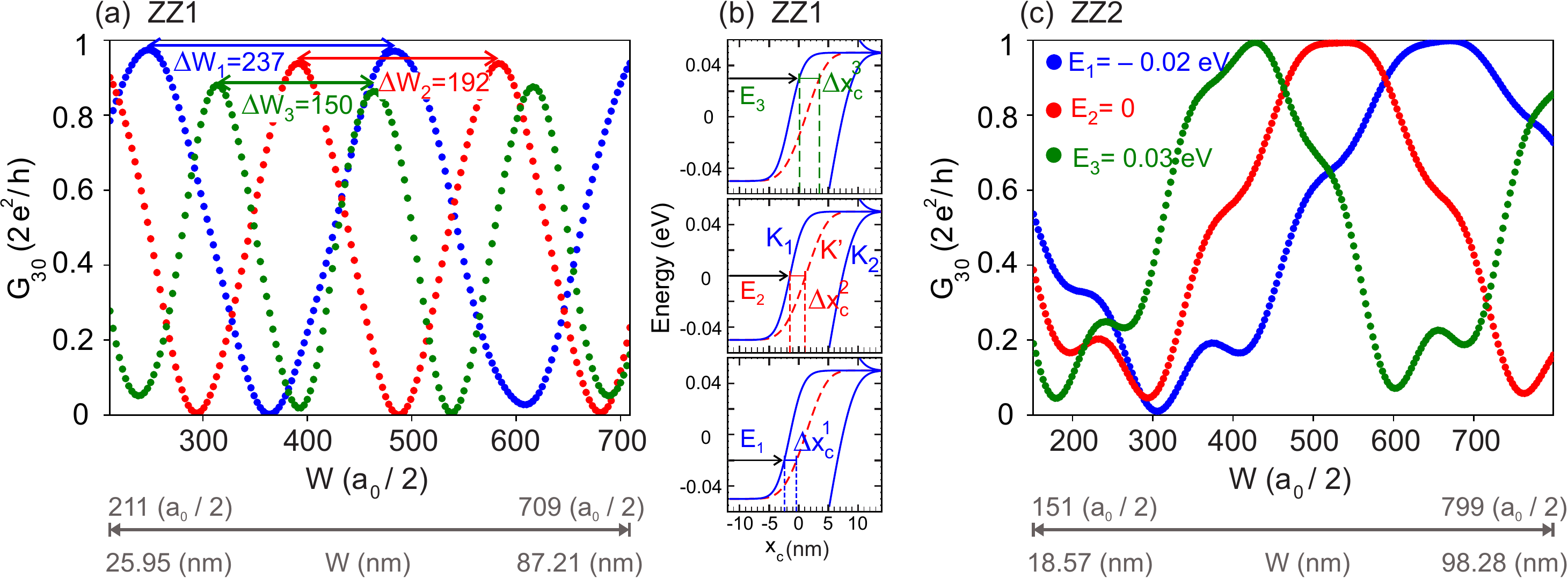}
	\caption{The Hall interface conductance $G_{30}$ of a bipolar MLG-BLG 
	junction as a function of ribbon width for (a) ZZ1 and (b) ZZ2 
	interfaces ($V_1 = -V_2 = -0.05 $ eV).
	The results are presented for three selected 
	Fermi energies $E_1 = -0.02$ eV (blue), $E_2 = 0$ (red), and $E_3 = 0.03$ eV
	(green).  
	Panel (b) is a zoom-in plot of the analytical energy levels of the ZZ1 interface 
	[shown in Fig.~\ref{fig3}(b)] around the bipolar energy window.
	In each plot, we show the spatial distance $\Delta x^i_c$ ($i=1,2,3$) between 
	the two neighboring edge channels $K_1$ and $K'$ corresponding to the 
	energy states $E_i$. 
	}
\label{fig5}
\end{figure*}

\subsection{Bipolar MLG-BLG quantum Hall bar} \label{bipo}

The character of quantum Hall edge transport in the \np\ junction is quite
different.
In Fig.\ \ref{fig3}, we plot the conductances for the same Hall bars 
as in the previous section, by applying the
potentials $V_1 = -V_0$ and $V_2 = V_0$ ($V_0 = 0.05$ eV) to the MLG and BLG
regions, respectively.
In this case, either side of the junction
gives electron- and hole-like edge modes with opposite edge chiralities and
results in metallic channels at the interface \cite{Abanin2007,Williams2007}. 
Accordingly, as seen in Figs.\ \ref{fig3}(e) and \ref{fig3}(f), the LIC 
within the bipolar energy $|E| < V_0$ is $G_{10} = 0$, whereas the Hall 
conductances $G_{20}$ and $G_{30}$ become nonzero and exhibit oscillatory 
behavior due to the presence of the interface channel.
Below, we discuss and show that these oscillations result from the interference
between two parallel edge states that belong to two different valleys. 
Note that the conductances for the ZZ1 and ZZ2 interfaces exhibit different profiles.
As seen, the Hall bar with ZZ1 interface exhibits more oscillatory behavior 
within the bipolar regime.
Further, in this regime, $G_{20}$, $G_{30}$ and $G_{31}$, $G_{30}$, 
respectively, satisfy the sum rules $G_{20} + G_{30} = 2e^2/h$ 
and $G_{31} + G_{30} = 4e^2/h$, reflecting the conservation 
of Hall edge modes in the MLG and BLG parts, respectively.
Therefore, only the HIC conductance $G_{30}$ is discussed hereafter.

To further characterize the transport fingerprints of the interface channels, 
we plot, in Figs.\ \ref{fig3}(b) and \ref{fig3}(d), the analytical energy levels 
of the MLG-BlG \np\ junction as a function of $x_c$ for the two interfaces, 
ZZ1 and ZZ2, respectively.
As seen, there are three edge channels near the MLG-BLG junction.
In both interfaces, two edge states belonging to two different valleys $K$ and $K'$,
and originating from zero electron and hole LLs are formed near the MLG-BLG 
junction (neighboring edge states), while the third one is formed rather far
away from the physical MLG-BLG boundary.
For the ZZ1 and ZZ2 interfaces, we will refer to them as ($K_1, K', K_2$) and 
($K, K'_1, K'_2$), respectively, starting from the left side in Figs.~\ref{fig3}(b) 
and \ref{fig3}(d).
Notice that these edge states are still spin degenerate.
Due to the interference between the spatially-separated edge states propagating 
along the \np\ junction, the cross-junction transport shows an oscillatory
behavior.
The coupling between the edge channels propagating along the \np\ junction
is illustrated schematically in Fig.~\ref{fig4}.
Two copropagating QH edge states encircle an enclosed area $S$ ($=Wd$), 
and thus, under the perpendicular magnetic field $B$, they acquire a phase 
difference $\Phi = B S$ arising from the AB effect.
The conductance oscillations can be described phenomenologically by
\cite{Wei2017,Jo2021}
\begin{equation} \label{eq:con-aha}
	G(E) \propto \cos(2 \pi \frac{\Phi}{\Phi_0} + \varphi_0),
\end{equation}
where $\varphi_0$ generally is an unknown phase associated with other effects.
However, here, because of our well-defined armchair-edged ribbons, it corresponds
to the angle between the valley isospins at the two edges of the nanoribbon
\cite{Tworzydlo2007}.
We argue and provide details below that the observed oscillations for 
the studied structures result from the AB interference between the QH edge 
states near the MLG-BLG \np\ junction.

\begin{figure}
	\centering
	\includegraphics[width = 8.5 cm]{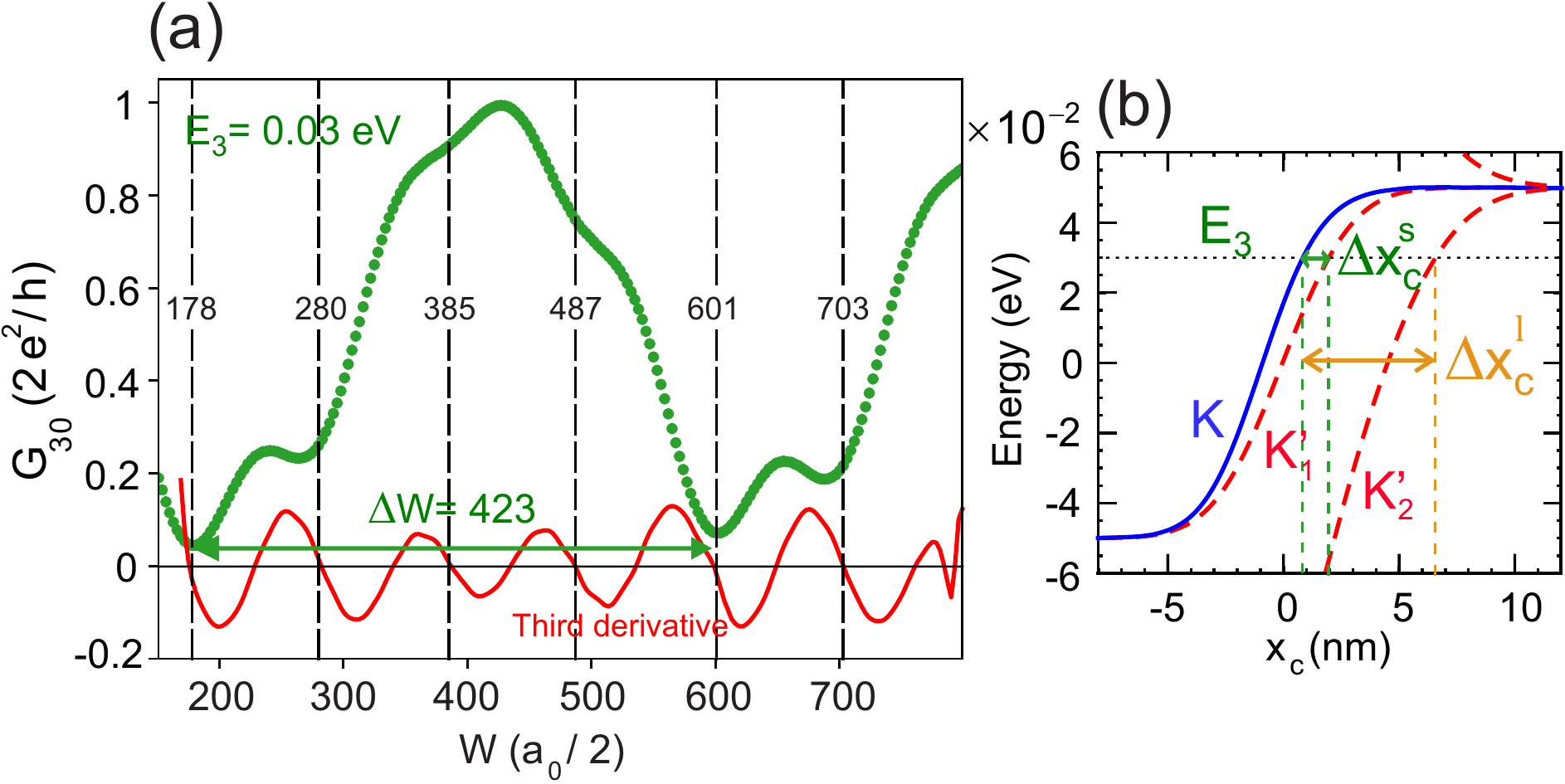}
	\caption{(a) The Hall interface conductance $G_{30}$ (green) and its third 
	numerical derivative (red) as a function of ribbon width for the ZZ2 
	interface at $E_3 = 0.03$ eV.
	Vertical dashed lines with the integer values of $W (a_0/2)$  indicate the 
	successive (local) minima positions. 
	(b) Zoom-in plot of the analytical energy levels of the ZZ1 interface 
	[shown in Fig.~\ref{fig3}(d)] around the bipolar energy window.
	Vertical green (orange) dashed lines show the $\Delta x_c^s$ ($\Delta x_c^l$) 
	as the distance between the two edge channels $K$ and $K'_1$ ($K$ and $K'_2$)
	at $E_3 = 0.03$ eV.
	}
\label{fig6}
\end{figure}

Figure \ref{fig5} shows the HIC conductance as a function of the 
Hall bar width for both interfaces (a) ZZ1 and (c) ZZ2. 
Here, we vary the width of the ribbons so that both remain metallic.
The results are presented for three representative Fermi energies in the bipolar 
regime, e.g., $E_1 = -0.02$ eV, $E_2 = 0$, and $E_3 = 0.03$ eV at $B = 60$ T.
In the case of the ZZ1 interface, shown in Fig.\ \ref{fig5}(a), we see regular
conductance oscillations with different periods for each Fermi energy as $W$ varies.
This implies that the variation of the AB phase $\Phi = B d \Delta W$
only comes from $\Delta W$, and that the spatial distance between the edge states 
($d$) along the \np\ junction is constant for each energy state.
From the AB phase, for a period of conductance oscillation in Fig.\ \ref{fig5}(a),
we obtain the spatial separation of the two edge channels
\begin{equation} \label{eq:delx}
	d = \frac{\Phi_0}{B \Delta W }
\end{equation}
for each energy state as $d_1 \approx 2.36$ nm, $d_2 \approx 2.92$ nm, and
$d_3 \approx 3.74$ nm.
Using the analytical energy spectrum [Fig.\ \ref{fig5}(b)], we also find 
$\Delta x_c^1 \approx 2.34$ nm, $\Delta x_c^2 \approx 2.93$ nm, and 
$\Delta x_c^3 \approx 3.70$ nm as the analytical splitting between the
two neighboring edge channels $K_1$ 
and $K'$, for the energy states $E_1$, $E_2$, and $E_3$, respectively.
Surprisingly, we find a strong agreement between the two results.

\begin{figure*}
	\centering
	\includegraphics[width = 17 cm]{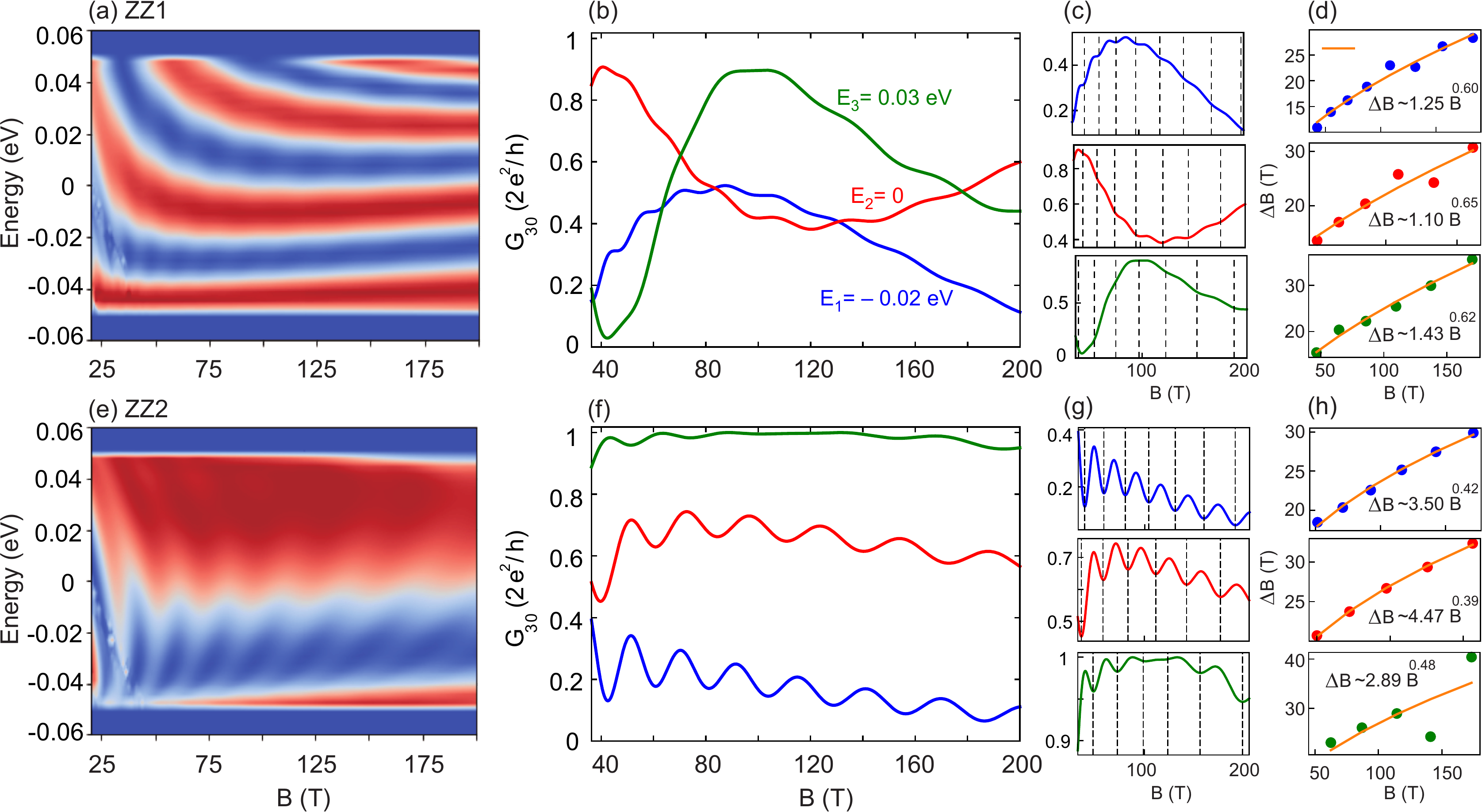}
	\caption{(Upper panels) (a) 2D color plot of the Hall conductance $G_{30}$ 
	of an \np\ MLG-BLG junction as functions of energy and magnetic field $B$ for 
	ZZ1 interface.
	(b) $G_{30}$ as a function of magnetic field for three selected energy states
	$E_1 = -0.02$ eV (blue), $E_2 = 0$ (red), and $E_3 = 0.03$ eV (green).
	Plots in (c) show the conductances separately for each energy state,
	with vertical dashed lines indicating successive minima positions.  
	(d) Corresponding magnetic field spacing ($\Delta B$) between the successive 
	minima extracted from panel (c).
	In each case, the fitting (orange curve) appears to show a $\sim B^{1/2}$ 
	dependence of $\Delta B $ as a function of magnetic field.
	(Lower panels) The same as the upper panels, but for the ZZ2 interface.
	}
\label{fig7}
\end{figure*}

Conductance behaves differently in the case of the ZZ2 interface. 
Whereas conductance for the ZZ1 interface shows only one set of oscillations, 
the ZZ2 interface shows two sets.
A double set of oscillations may indicate the presence of two distinct AB 
interference loops operating near the \np\ junction.
We attribute the small oscillations to the AB interference between the 
$K$ and $K'_2$ edge channels and the large ones to the AB interference between 
the two neighboring edge states, i.e., $K$ and $K'_1$ [see Fig.~\ref{fig6}(b)].
To support our statement quantitatively, the conductance corresponding to the energy
state of, e.g., $E_3 = 0.03$ eV and its numerical third derivative as a 
function of $W$ are shown in Fig.\ \ref{fig6}(a).
Using the Eq.\ \eqref{eq:delx}, we find (averaged) $d^{\mathrm l} = 5.35$ nm 
for small $\Delta W$ periods, which are represented by vertical dashed lines 
in Fig.~\ref{fig6}(a).
The corresponding spatial distance between the $K$ and $K'_2$ edge states, 
extracted from the analytical results [shown in orange in Fig.\ \ref{fig6}(b)], 
is $\Delta x_c^{\mathrm l} = 5.47$ nm, which agrees very well with the $d^l$ 
obtained from the AB-interference description.
Using Eq.~\eqref{eq:delx} with $\Delta W = 423\, a_0 / 2$ between two minima of the 
large oscillation [Fig.\ \ref{fig6}(a)], we obtain $d^{\mathrm s} = 1.32$ nm, 
which agrees well with $\Delta x_c^{\mathrm s} = 1.29$ nm extracted from the 
analytical results for the two neighboring edge states ($K$, $K'_1$) at the
corresponding energy state $E_3 = 0.03$ eV, as shown by the green arrow 
in Fig.\ \ref{fig6}(b).
The two measurements are perfectly consistent and confirm our interpretation.

Notice that in both interfaces, conductance oscillation amplitude $\Delta G_{03}$
as a result of coupling between the two adjoining edge channels 
(belonging to distinct valleys), varies approximately between 0 and $2e^2/h$.
$\Delta G_{03}$ variation owing to the coupling between two far-distant edge 
channels, on the other hand, is not noticeable in the case of the ZZ2 interface 
and is nearly absent in the ZZ1 interface.
This indicates that the AB interference between the two edge channels with 
opposite valleys and spatially adjacent to each other is significant and mostly
mediates the transport across the junction.

Probability densities corresponding to examples of energies, as labeled 
by $E_{1,2}$ in Figs.~\ref{fig3}(e) and \ref{fig3}(f), are shown at the right 
of each panel.
A (valley-degenerate) edge channel coming from lead 0 in 
the MLG region is wholly
guided along the \np\ junction at the intersection between the ribbon 
physical edge and the \np\ interface, and splits into valley-polarized channels 
due to the presence of the second layer of BLG region, which, after 
traveling along the \np\ junction, obtain different AB phases, interfere
at the bottom ribbon physical edge and result in two complementary edge channels,
collected by leads 1 and 2.
As can be seen, the outcome leads to full ($E_1$ in each panel) or finite 
($E_2$) transmission, depending on the Fermi energy. 
This mechanism is the electronic analogue of the optical MZ interferometer, which 
was first introduced by Ji \textit{et al.}\ \cite{Ji2003}. 

\begin{figure*}
	\centering
	\includegraphics[width = 17 cm]{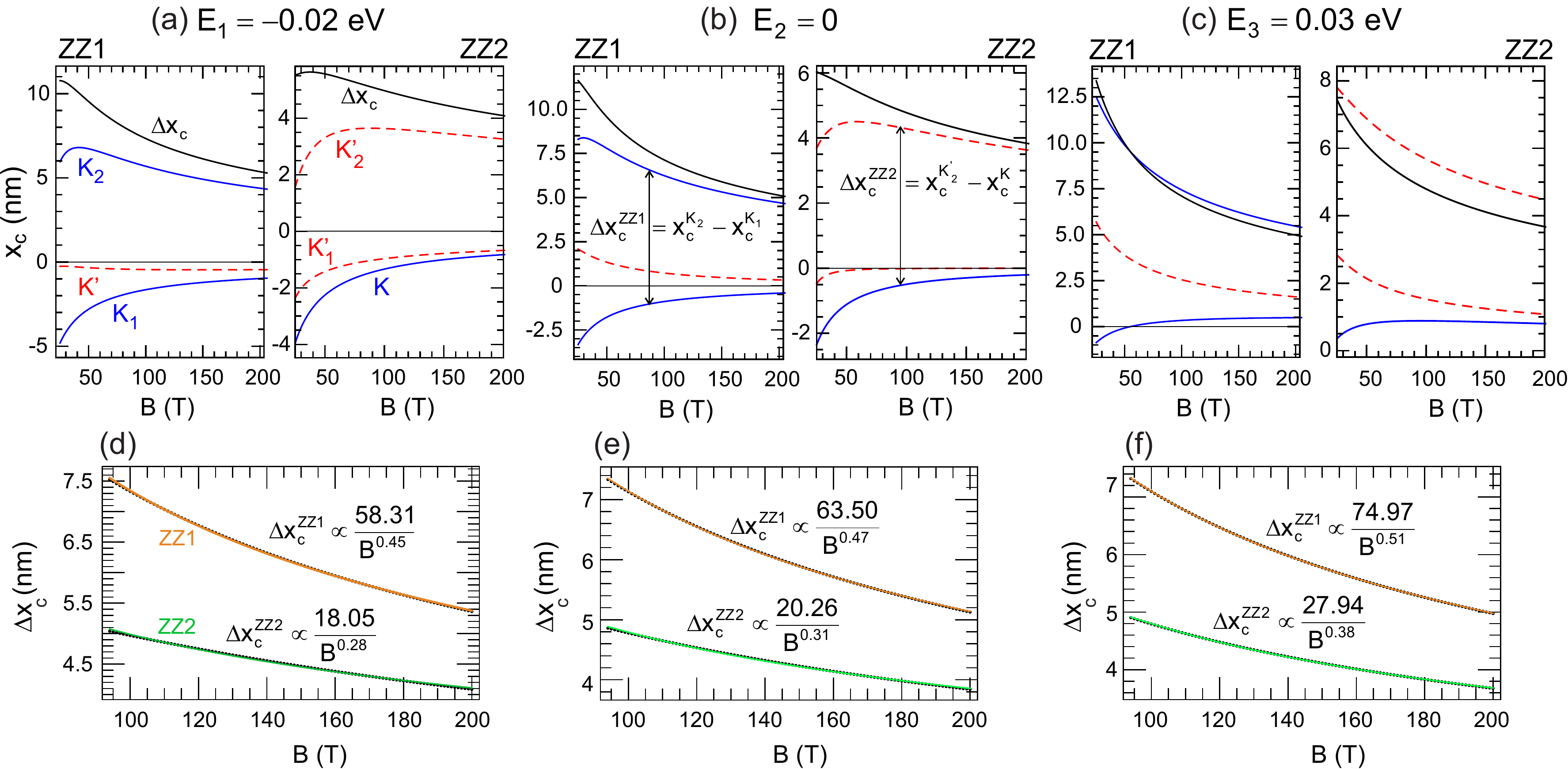}
	\caption{The cyclotron orbit position $x_c$ of the three edge channels 
	formed in the bipolar regime 
	[Figs~\ref{fig3}(b) and \ref{fig3}(d)] as a function of the magnetic field
	for three examples of the energy states (a) $E_1 = -0.02$ eV, 
	(b) $E_2 = 0$, and (c) $E_3 = 0.03$ eV.
	In each case, the left (right) panel shows the result for ZZ1 (ZZ2) interface.
	Blue solid (red dashed) curves refer to the $K$ ($K'$) valley.
	Black curves depict the difference between the two far-distant edge channels 
	which is defined as $\Delta x_c^{ZZ1} = |x_c^{K_2} - x_c^{K_1}|$ and
	$\Delta x_c^{ZZ2} = |x_c^{K'_2} - x_c^{K}|$ for the ZZ1 and ZZ2 intefaces,
	respectively.
	The lower panels show the power-law dependence 
	$\sim \alpha B^\beta $ of
	the $\Delta x_c$ for the ZZ1 (orange) and ZZ2 (green) interfaces corresponding
	to the upper panels' energies (d) $E_1$, (e) $E_2$, and (f) $E_3$.
	}
\label{fig8}
\end{figure*}

Now, we analyze the magnetic-field dependence of the HIC conductance for both
interfaces.
Figures \ref{fig7}(a) and \ref{fig7}(e) show a two-dimensional (2D) color plot of 
HIC as a function of magnetic field in the bipolar energy window for both 
interfaces ZZ1 and ZZ2, respectively.
One can obviously see two different conductance profiles for two boundaries.
While the conductance as a function of energy in the ZZ1 interface becomes more
oscillatory as $B$ increases, the ZZ2 interface exhibits low and high conductances 
for low and high energies, respectively.
In the following, we show that the $B$-dependency of the HIC in both
cases supports the AB oscillations idea for the considered MLG-BLG interfaces.

Figures \ref{fig7}(b) and \ref{fig7}(f) show the magnetoconductance at three 
selective Fermi energies
$E_1 = -0.02$ eV, $E_2 = 0$, and $E_3 = 0.03$ eV for the ZZ1 and ZZ2 interfaces, 
respectively.
The magnetoconductance oscillations are reminiscent of the AB oscillation. 
Here also notice that a double set of oscillations can be considered for both
interfaces.
Small oscillations (due to AB interference between two distant edge channels), 
as seen in Figs.~\ref{fig7}(b) and \ref{fig7}(f), and large oscillations 
(due to coupling between two neighboring edge channels), whose periodicities 
are not covered in the shown $B$-range.
For both interfaces, in the case of small oscillations, all three 
conductances reveal a different period of $\Delta B$, as expected.  
Because, as previously stated, the spatial separation ($d$) of the edge channels 
varies depending on the energy state.
According to the relation $\Delta B (W d) = \Phi_0$, one expects a constant 
$\Delta B$ for a fixed edge-channel separation $d$ at each energy state.
However, magnetoconductances exhibit a common trend, namely the 
increase of $\Delta B$ with increasing of $B$ [Figs.~\ref{fig7}(c,d,g,h)].
The magnetic-field spacing ($\Delta B$) derived from the successive maxima 
(or minima) of the magnetoconductance curves is separately shown for each 
energy state in Figs.~\ref{fig7}(d) and \ref{fig7}(h) for both interfaces.
It seems that both interfaces share a common trend $\sim B^{1/2}$.
Our calculations for a wide range of $B$ fields, such as $40-500$ T (not shown here), 
show this dependency with high precision, i.e., $\Delta B \propto B^{0.50}$ 
and consequently the spatial separation of the edge channels decreases by 
$d \propto 1 / B^{1/2}$.
This behavior is in contrast with that of an MLG \np~junction for which the
magnetoconductance oscillations in the QH regime reveal a linear  decrease 
of $\Delta B$ as a function of $B$ \cite{Morikawa2015,Makk2018}.

It is also worth noting that the (small) oscillation amplitude for the 
ZZ1 boundary is not remarkable and vanishes as the magnetic field 
increases; for the ZZ2 boundary, it persists for the entire magnetic-field 
range, indicating that coupling between two far-distant edge channels in the ZZ1 
boundary is weaker than that of the ZZ2 boundary which is consistent with what 
we discussed in Figs.~\ref{fig5}(a) and \ref{fig5}(c) where the ZZ2 boundary 
shows a double set of AB interference.
This can be understood as follows.
As shown in Fig.~\ref{fig4},
the two rightmost edge channels of the ZZ1 boundary are from two different 
valleys ($K',K_2$), whereas the two rightmost edge channels of the ZZ2 boundary 
are from the same valley ($K'_1,K'_2$).
As a result, any interchannel scattering from the leftmost edge channel 
(belonging to $K$ valley in both cases) to the rightmost channel in the case of 
ZZ1 interface requires two intervalley scattering process 
$K \rightarrow K' \rightarrow K$ costing more energy than the ZZ2 case, 
which requires only one scattering process $K \rightarrow K'$.  

The above-predicted $B$-dependence of the spatial separation of the edge channels,
i.e., $d \propto 1 / B^{1/2}$, is consistent with our analytical results.
Figure \ref{fig8} depicts the cyclotron orbit position $x_c$
of the (three) interface edge channels [Figs~\ref{fig3}(b) and \ref{fig3}(d)]
as a function of the magnetic field
for the studied selective energies (a) $E_{1} = -0.02$ eV, (b) $E_{2} = 0$, 
and (c) $E_{3} = 0.03$ eV.
In each case, the results are presented for both interfaces, ZZ1 (left panel) and 
ZZ2 (right panel).
We also plot the $\Delta x_c$ (black curves) for the two far-distant edge channels
defined in the ZZ1 and ZZ2 plots shown in Fig.~\ref{fig8}(b).
As seen, in both cases, the spatial separation of the edge channels decreases 
as the $B$ field increases.
By fitting the $\Delta x_c$ to the power low $\alpha / B^\beta$ 
[lower panels in Fig.~\ref{fig8}], 
we obtain a $\sim 1/B^{0.50}$ and $\sim 1/B^{0.30}$ dependence for the edge-channel
separations in the case of ZZ1 and ZZ2 interfaces, respectively.
This is in agreement with what was predicted in the magnetoconductance oscillations
shown in Figs.~\ref{fig7}(d) and \ref{fig7}(h). 

\begin{figure}
	\centering
	\includegraphics[width = 8.5 cm]{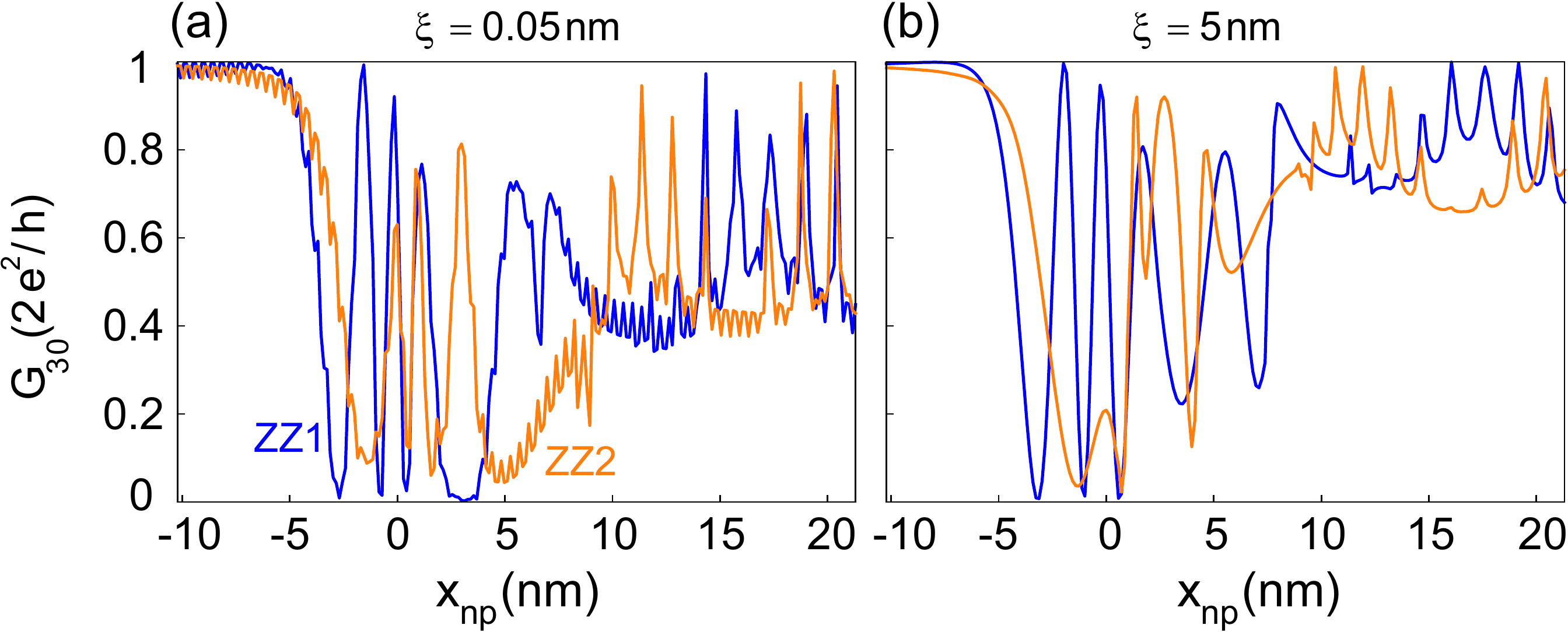}
	\caption{The Hall interface conductance $G_{30}$ of a bipolar MLG-BLG 
	junction as a function of the \np-junction position $x_{np}$ at
	Fermi energy $E = 0$ for ZZ1 (blue) and ZZ2 (orange) interfaces when the applied
	potential is (a) abrupt ($\xi = 0.05$~nm) and (b) smoothly varying ($\xi = 5$~nm).  
	}
\label{fig9}
\end{figure}

To complement our analysis, we perform the transport calculations for
the structures when the \np-junction position 
does not coincide with the physical MLG-BLG interface at $x_0 = 0$.
Figure \ref{fig9}(a) shows the HIC of a bipolar MLG-BLG junction as a function 
of the \np-junction position, $x_{np}$, at Fermi energy $E = 0$ for both interfaces.  
The results are presented in steps of $a_{cc} = 0.142$~nm (C-C distance) from 
$x_{np} = -72 a_{cc} \approx -10.23$~nm in the MLG region to 
$x_{np} = 150 a_{cc} \approx 21.30$~nm in the BLG region.
The small rapid oscillations in Fig.\ \ref{fig9}(a) are caused by an abrupt step 
potential ($\xi = 0.05$ nm), which does not exist with a smoothly varying 
potential ($\xi = 5$ nm), as shown in Fig.~\ref{fig9}(b).
We see that when the \np\ junction ($x_{np}$) is tuned in the MLG region 
far away from the MLG-BLG interface, the conductance shows a plateau at 
$2e^2/h$ for both interfaces, which is consistent with previous study in
Ref.~\cite{Tworzydlo2007}.
This result is also consistent with our analytical calculations, 
which show that in the case of the MLG \np\ junction, the formation of an edge 
channel in the bipolar energy is not valley polarized; 
see Fig.~\ref{figA1}(a) in the appendix.
As a result, there is no AB interference between the two degenerate edge channels
propagating along the \np\ interface, resulting in a conductance plateau 
at $2e^2/h$ \cite{Tworzydlo2007}.
However, as previously stated, valley- and spin-polarized edge channels in 
a graphene \np\ junction can be created experimentally, allowing for the 
realization of a MZ interferometer in such structures 
\cite{Morikawa2015,Wei2017,Makk2018,Jo2021}.

Approaching the $x_{np}$ to the physical MLG-BLG interface, where the valley
degeneracy is lifted for both interfaces, 
influences the HIC as a result of the AB interference.
Notice that the two interfaces exhibit different profiles.
While the ZZ1 interface exhibits full transmissions or full back-reflections 
depending on the exact position of the $x_{np}$ in the MLG region, 
the ZZ2 interface exhibits a finite transmission that approaches zero 
as the $x_{np}$ approaches to the MLG-BLG junction.
This property is shared by all energies in the bipolar regime.
Shallow oscillations persist in the BLG region in consistent with our 
analytical calculations, which show two (valley-degenerate) edge channels along 
the \np\ junction of a gated BLG structure in the bipolar regime
[see Fig.~\ref{figA1}(b)].


\section{Conclusion} \label{con}
In conclusion, we demonstrated that an \np\ junction of MLG-BLG interface bar in 
the Hall regime results in valley-polarized edge-channel interferences and can 
function as a fully tunable MZ interferometer device.
Using the Landauer-B\"uttiker formalism along with the TBM,
we investigated the conductance properties of unipolar and bipolar hybrid MLG-BLG
junctions in two different interfaces known as ZZ1 and ZZ2 boundaries. 
Our findings show that, in contrast to the ZZ2 interface, the ZZ1 interface affects 
the higher QH plateaus in a unipolar MLG-BLG junction, indicating that the coupling 
between the MLG and BLG regions is weaker in the case of the ZZ1 interface. 
Furthermore, no HIC was observed in either type of MLG-BLG junction in 
the unipolar regime.

In the bipolar regime, we found that both types of MLG-BLG interfaces exhibit 
a gate tunable HIC with resonant behavior as a function of Fermi energy, 
which is different for each interface.
By investigating the bar-width dependence of the conductance oscillations and 
solving the Dirac-Weyl equation analytically for a gated semi-infinite MLG-BLG 
junction, we demonstrated that the conductance oscillations result from AB 
interference between three valley-polarized (but spin degenerate) Hall edge 
channels propagating along the MLG-BLG \np\ junction.
We found that the coupling between the two neighboring opposite-valley-polarized 
edge channels is predominant and results in a large conductance oscillation.
By investigating the magnetic-field dependence of the conductance oscillations, 
we found a small-amplitude oscillation for both interfaces, resulting from the 
AB interference between the two far-distant edge channels.
The small oscillation in the ZZ1 boundary is not noticeable and disappears when 
the magnetic field is increased; however, it persists in the ZZ2 boundary for a 
long period of magnetic field.

Finally, while realistic samples of such hybrid structures would be more complex 
than those modeled here, we believe that the main features of our 
results can be captured in relevant experimental systems.
Such a \textit{natural} junction between MLG and BLG in QH regime 
can be a promising platform to study electron interference associated with 
valley-polarized edge channels.
Two possible areas of electron-interferometry research are fractional and 
non-Abelian statistics \cite{Law2006,Feldman2006,Bid2009} and 
quantum entanglement via two-particle interference
\cite{Yurke1992,Samuelsson2004}.

\section*{Acknowledgements}
This work was supported by the Institute
for Basic Science in Korea (No. IBS-R024-D1).

\renewcommand{\theequation}{A\arabic{equation}}
\setcounter{equation}{0}
\renewcommand{\thefigure}{A\arabic{figure}}
\setcounter{figure}{0}

\section*{Appendix} \label{append}

In this Appendix, we briefly review the main steps of our analytical calculations.
For more details see Refs.\ \cite{Koshino2010,Mirzakhani2017MBM}.
In the presence of a perpendicular magnetic field $\mb B = B \ez$ the dynamics of
the carriers in MLG is described by the Dirac-Weyl Hamiltonian (for the $K$ valley)
 \cite{Beenakker2008,Koshino2010} 
\begin{equation} \label{eqn:H_mono}
  \mc{H}^{K}_\mathrm{MLG} =
   \Big(\begin{array}{cc}
       V_1 & v_F \pi_{-} \\
       v_F \pi_{+} & V_1 \\
   \end{array}\Big),
\end{equation}
where $ \pi_{\pm} = \pi_{x} \pm \i \pi_{y} $ with 
$\mbox{\boldmath$\pi$} = -\i \hbar \mbox{\boldmath$\nabla$} + e\mb{A}$ 
the kinetic momentum operator, 
$ \mb{A} = (0,Bx,0) $ is the vector potential in the Landau gauge, 
and $V_1$ the potential applied to MLG.
The Hamiltonian at the $K'$ valley is obtained by interchanging 
$\pi_+$ with $\pi_-$ in Eq.\ \eqref{eqn:H_mono}.
In terms of the TB parameters, the Fermi velocity is defined as 
$v_F = 3 \gamma_0 a_{cc} / 2\hbar \approx 8.74 \times 10^5$ m/s, 
where $\gamma_0 = 2.7$ eV is the nearest-neighbor 
intralayer hopping parameter and $a_{cc} = 0.142$ nm the C-C distance in 
graphene hexagon.

In the Landau gauge, the Hamiltonian \eqref{eqn:H_mono} is translationally 
invariant in the $y$ direction and its two-component eigenstates take the form
$ \mb{\Psi}^{K}(\mb{r}) = [\psi^{K}_{A}(x), \psi^{K}_{B}(x)]^T \e^{\i k_y y} $,
where $\psi^K_A$ and $\psi^K_B$ are the envelope functions on the  
sublattices $A$ and $B$, respectively. 

Applying the Schr\"{o}dinger equation for the two-component envelope function 
$\mb{\Psi}^K (\mb{r})$ 
\begin{equation} \label{eqn:Schr}
   \mathcal{H}^{K} \mb{\Psi}^{K}(\mb{r}) = E~\mb{\Psi}^{K}(\mb{r}),
\end{equation}
and doing some algebra, we obtain
\begin{equation} \label{eqn:pi}
   v_F \pi_{+} = \i E_0 b^{\dag},\quad
   v_F \pi_{-} = -\i E_0 b,
\end{equation}
where $ E_0 = \sqrt{2} \hbar v_F /l_{B}$ and $l_{B}= \sqrt{\hbar/eB}$ is 
the magnetic length. 
We have also introduced the raising and lowering operators
\begin{equation} \label{eqn:def a}
  b^\dag = -\p_z + z/2, \quad
   b = \p_z + z/2,
\end{equation}
where the dimensionless coordinate $z$ is defined by
\begin{equation} \label{eqn:def z}
   z = \sqrt{2}(x-x_c)/l_{B},
\end{equation}
and $x_c=-k_{y}l^{2}_{B}$ is the center of the cyclotron orbit.

Decoupling the Schr\"{o}dinger equation for the spinor components 
of the envelope function $\mb{\Psi}^K (\mb{r})$
leads to the Weber differential equation \cite{Gradshteyn}
\begin{equation} \label{eqn:F_K_B}
  (\nu-b^{\dag}b)\,\psi^{K}_{B} = \Big(\frac{\partial^{2}}{\partial z^{2}}
  + \nu + \frac{1}{2} - \frac{z^{2}}{4}\Big) \psi^{K}_{B}(z) = 0,
\end{equation}
with
\begin{equation} \label{eqn:nu mono}
   \nu = (\varepsilon-v_1)^{2}.
\end{equation}
Here $\varepsilon = E/E_0$ and $v_1= V_1/ E_0$.
\begin{figure}
	\centering
	\includegraphics[width = 7.5 cm]{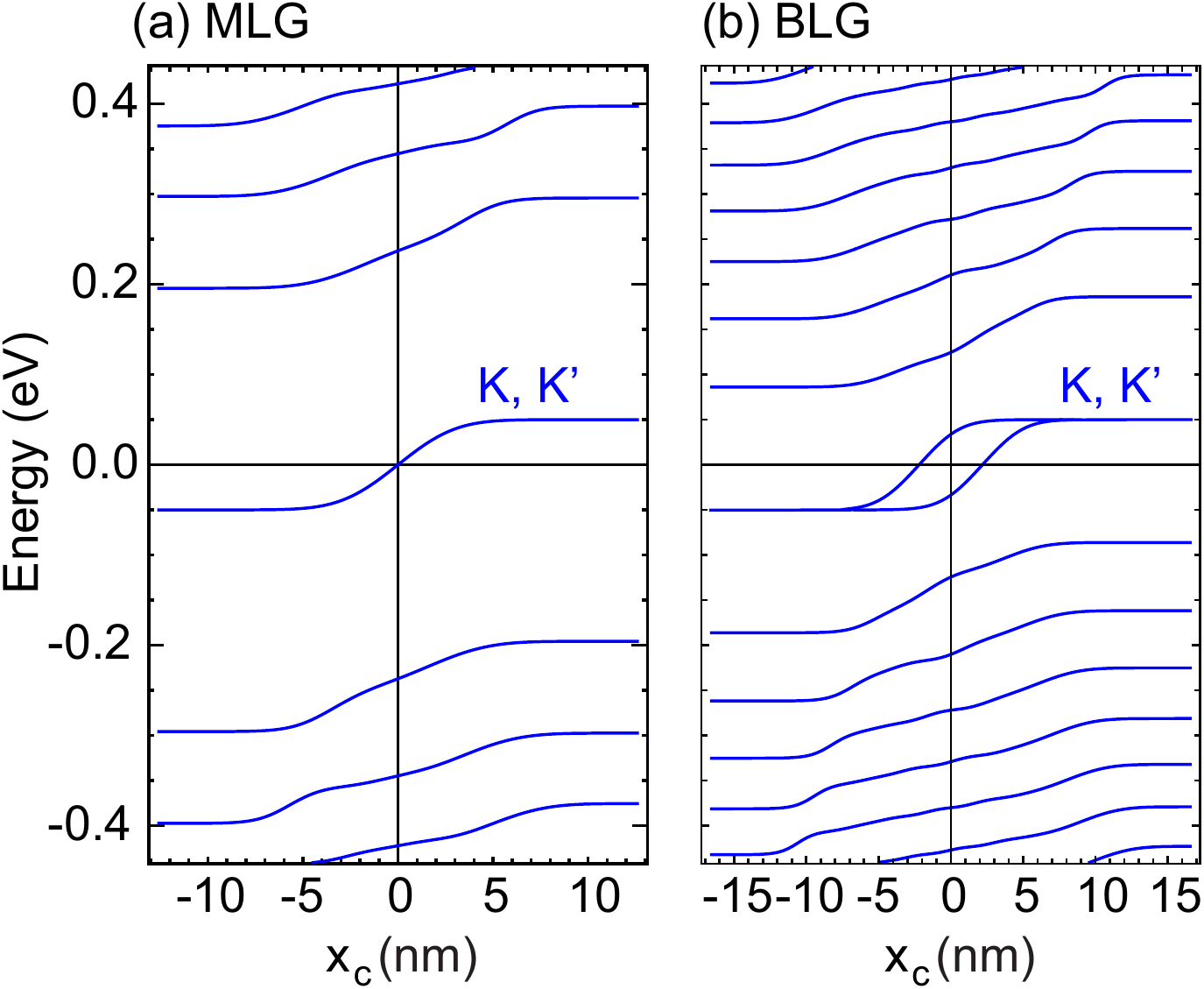}
	\caption{Energy levels of an \np\ junction of a pristine (a) MLG and (b) BLG 
	as a function of the center of 
	the cyclotron orbit $x_c$ obtained from analytical results.
	}
\label{figA1}
\end{figure}

The two independent solutions of Eq.\ \eqref{eqn:F_K_B} are the 
\textit{parabolic cylinder functions} $D_\nu(z)$ and $D_{\nu}(-z)$ which 
vanish in the 
limit $z \rightarrow \infty$ and $z \rightarrow - \infty$, respectively.
The other spinor component $\psi^{K}_{A}(z)$ can be obtained from the
Schr\"{o}dinger equation and by employing the relations
\begin{align} \label{eqn:ident}
  & b^{\dag} D_{\nu}(z)= \mathrm{sgn}(z) D_{\nu+1}(z), \notag \\
  & b D_{\nu}(z)= \mathrm{sgn}(z) \nu D_{\nu-1}(z),
\end{align}
where $\mathrm{sgn}(z)$ is the sign function.
    
Thus the spinor components in the MLG region where $ x < 0 $ is given by 
\begin{equation}\label{eqn:MLGwave}
  \left(\begin{array}{c}
     \psi^{K}_{A}(x) \\
     \psi^{K}_{B}(x) \\
   \end{array}
   \right) =
   C \left(
    \begin{array}{c}
      \i (\varepsilon-v_1) D_{\nu-1}(-z) \\
      D_{\nu}(-z)
    \end{array}
    \right),
 \end{equation}    
with $C$ being the normalization constant.    
The wave function at the $K'$ valley can be obtained by
$ [\psi^{K'}_{A}, \psi^{K'}_{B}] = [\psi^{K}_{B}, \psi^{K}_{A}] $.

The BLG region can be described in terms of four sublattices, 
labeled $A1$, $B1$, for the lower layer and $A2$, $B2$,
for the upper layer. 
We only include the coupling between two atoms stacked on top of each 
other, i.e., $B1$ and $A2$ [see Fig.~\ref{fig1}], and ignore the small 
contributions of the other interlayer couplings.
In the vicinity of the $K$ valley, the effective Hamiltonian is
\cite{Koshino2010,McCann2006}
\begin{equation}\label{eqn:H_Bi} 
  \mc{H}^{K}_\mathrm{BLG} = \left(
    \begin{array}{cccc}
    V_2 & v_F \pi_{-} & 0 & 0 \\
    v_F \pi_{+} & V_2 & \gamma_{1} & 0 \\
    0 & \gamma_{1} & V_2 & v_F \pi_{-} \\
    0 & 0 & v_F \pi_{+} & V_2 \\
    \end{array}\right),
\end{equation}
where $\gamma_1$ is the nearest-neighbor interlayer hopping term and 
$V_2$ is the potential applied to the BLG region.

Solving the Schr\"{o}dinger equation \eqref{eqn:Schr} for the four-component 
envelope function 
$\mb{\Phi}^{K}(\mb{r}) = (\phi^K_{A1}, \phi^K_{B1}, \phi^K_{A2}, \phi^K_{B2})^T \e^{\i k_y y}$
and using  the relations \eqref{eqn:ident}
we obtain
\begin{align} \label{eqn:BLGenfun}
  \left(
   \begin{array}{c} 
     \phi^{K}_{A1}(x) \\
     \phi^{K}_{B1}(x) \\
     \phi^{K}_{A2}(x) \\
     \phi^{K}_{B2}(x) \\
     \end{array}\right) =
     \sum_{\mu = \pm}
     & C_{\mu} \left(
       \begin{array}{c}
       - \i [\tilde{\gamma}_1 \varepsilon' \nu_\mu / (\varepsilon'^2 - \nu_\mu)] 
       D_{\nu_{\mu} - 1}(z) \\
       \ [\tilde{\gamma}_1 \varepsilon'^2 /(\varepsilon'^2 - \nu_\mu)] 
       D_{\nu_{\mu}}(z) \\
       \varepsilon' D_{\nu_{\mu}}(z) \\
       \i D_{\nu_{\mu} + 1}(z)
       \end{array}
   \right),
\end{align}
where $C_{\mu}$ is normalization constant, $ \tilde{\gamma}_1 = \gamma_1/ E_0 $,
$ \varepsilon' = \varepsilon - v_2 $ with $ v_2 = V_2 / E_0 $, 
and
\begin{equation}\label{eqn:nubi}
   \nu_{\mu}=\frac{1}{2} \Big( - 1 + 2\varepsilon'^2 + 
   \mu \sqrt{1 + 4 \tilde{\gamma_{1}}^{2} \varepsilon'}\,\Big).
\end{equation}
The wave function at the $K'$ valley can be obtained by
$ [\psi^{K'}_{A1}, \psi^{K'}_{B1}, \psi^{K'}_{A2}, \psi^{K'}_{B2}] 
= [\psi^{K}_{B2}, \psi^{K}_{A2}, \psi^{K}_{B1}, \psi^{K}_{A1}] $.

The boundary conditions at $ x = 0  $ for each interface are \cite{Nakanishi2010}:
\begin{equation}\label{}
\text{ZZ1} 
\left \{\!\!
\begin{array}{l}
  \phi^\tau_{A1}(z_0)=\psi^\tau_A(z_0), \notag   \\
  \phi^\tau_{B1}(z_0)=\psi^\tau_B(z_0), \notag   \\
  \phi^\tau_{B2}(z_0)=0,  
\end{array} \right.
\quad
\text{ZZ2}
\left \{\!\!
\begin{array}{l}
  \phi^\tau_{A1}(z_0)=\psi^\tau_A(z_0), \notag   \\
  \phi^\tau_{B1}(z_0)=\psi^\tau_B(z_0), \notag   \\
  \phi^\tau_{A2}(z_0)=0,  
\end{array} \right.
\end{equation}
where $\tau=K$ or $K'$ and $z_0 = -\sqrt{2}~x_c/l_{B} \equiv z(x=0)$.

The above conditions for two interfaces in each valley lead to a system of 
equations from which the eigenvalues are obtained by setting the determinant 
of the coefficients to zero.
Solving such determinants numerically, one can obtain eigenvalues as a function 
of, e.g., cyclotron orbit $x_c$, as presented in Figs~\ref{fig2}(e,f) and 
\ref{fig3}(b,d).

Using the analytical results, we also plot the energy levels for the \np\ junctions of 
pure MLG and BLG structures in Figs.~\ref{figA1}(a) and \ref{figA1}(b), respectively.
As mentioned in the main text, one can clearly see one and two (valley degenerate) 
edge channels along the \np\ junction in the bipolar regime for MLG and BLG 
structures, respectively.
    

\end{document}